\documentstyle[lprocl,epsfig,11pt]{article}
\textheight
8.5in  \parskip 7pt \openup1\jot 
\def\be{\begin{equation}}
\def\ee{\end{equation}}
\def\bea{\begin{eqnarray}}
\def\eea{\end{eqnarray}}

\newskip\humongous \humongous=0pt plus 1000pt minus 1000pt
\def\caja{\mathsurround=0pt}
\def\eqalign#1{\,\vcenter{\openup1\jot \caja
        \ialign{\strut \hfil$\displaystyle{##}$&$
        \displaystyle{{}##}$\hfil\crcr#1\crcr}}\,}
\newif\ifdtup

\def\eqright #1\cr{\noalign{\hfill$\displaystyle{{}#1}$}}
\def\eqleft #1\cr{\noalign{\noindent$\displaystyle{{}#1}$\hfill}}

\def\oldreffmt#1{\rlap{[#1]} \hbox to 2\parindent{}}

\def\figfmt#1{\rlap{Figure {#1}} \hbox to 1in{}}

\def\begineq #1\endeq{$$ \refstepcounter{equation}\eqalign{#1}\eqno
	(\theequation) $$}
\def\contlimit{\,{\hbox{$\longrightarrow$}\kern-1.8em\lower1ex
\hbox{${\scriptstyle (a\rightarrow0)}$}}\,}
\def\centeron#1#2{{\setbox0=\hbox{#1}\setbox1=\hbox{#2}\ifdim
\wd1>\wd0\kern.5\wd1\kern-.5\wd0\fi
\copy0\kern-.5\wd0\kern-.5\wd1\copy1\ifdim\wd0>\wd1
\kern.5\wd0\kern-.5\wd1\fi}}
\def\centerover#1#2{\centeron{#1}{\setbox0=\hbox{#1}\setbox
1=\hbox{#2}\raise\ht0\hbox{\raise\dp1\hbox{\copy1}}}}
\def\centerunder#1#2{\centeron{#1}{\setbox0=\hbox{#1}\setbox
1=\hbox{#2}\lower\dp0\hbox{\lower\ht1\hbox{\copy1}}}}
\def\lsim{\;\centeron{\raise.35ex\hbox{$<$}}{\lower.65ex\hbox
{$\sim$}}\;}
\def\gsim{\;\centeron{\raise.35ex\hbox{$>$}}{\lower.65ex\hbox
{$\sim$}}\;}

\def\super#1{\ifmmode \hbox{\textsuper{#1}}\else\textsuper{#1}\fi}
\def\textsuper#1{\newcount\holdspacefactor\holdspacefactor=\spacefactor
$^{#1}$\spacefactor=\holdspacefactor}

\def\getcite#1,{\advance\citenumber by1
\ifnum\citenumber=1
\ref{#1}\let\next=\getcite\else\ifx#1@\let\next=\relax
\else ,\ref{#1}\let\next=\getcite\fi\fi\next}

\def\upon #1/#2 {{\textstyle{#1\over #2}}}
\relax

\def\til#1{\centeron{\hbox{$#1$}}{\lower 2ex\hbox{$\char'176$}}}
\def\tild#1{\centeron{\hbox{$\,#1$}}{\lower 2.5ex\hbox{$\char'176$}}}
\def\sumtil{\centeron{\hbox{$\displaystyle\sum$}}{\lower
-1.5ex\hbox{$\widetilde{\phantom{xx}}$}}}

\def\pom{{\rm P\kern -0.53em\llap I\,}}
\def\spom{{\rm P\kern -0.36em\llap \small I\,}}
\def\sspom{{\rm P\kern -0.33em\llap \footnotesize I\,}}

\begin{document} 

\begin{titlepage} 

\rightline{\vbox{\halign{&#\hfil\cr
&ANL-HEP-CP-97-73 \cr
&\today\cr}}} 
\vspace{1.25in} 

\begin{center}
{\bf THE VIIth BLOIS WORKSHOP: THEORY SUMMARY AND FACTORIZATION ISSUES 
}\footnote{Work 
supported by the U.S.
Department of Energy, Division of High Energy Physics, \newline Contracts
W-31-109-ENG-38 and DEFG05-86-ER-40272} 
\medskip

Alan. R. White\footnote{arw@hep.anl.gov }
\end{center}
\vskip 0.6cm

\centerline{High Energy Physics Division}
\centerline{Argonne National Laboratory}
\centerline{9700 South Cass, Il 60439, USA.}
\vspace{0.5cm}

\begin{abstract} 

Workshop presentations on elastic and diffractive scattering and other recent 
advances in hadron physics are summarized. The role of ``factorization''
in determining parton properties of the pomeron is particularly discussed.

\end{abstract} 

\vspace{2.5in}
\begin{center}

Presented at the International Conference (VIIth Blois Workshop) on Elastic
and Diffractive Scattering - Recent Advances in Hadron Physics.
\newline  Seoul, Korea. June 10-14, 1997.

\end{center}

\end{titlepage}

\section{Introduction}

Theory presentations at this workshop have covered a wide range of topics. 
In addition to the traditional topics of elastic and diffractive scattering,
we have had a variety of interesting talks coming under the broad umbrella 
of ``Recent Advances in Hadron Physics''. These have included 
review talks on lattice gauge theory, techniques for high-order 
perturbative QCD calculations, strong interaction effective field theories,
the current status of QED and the construction of theories beyond the
Standard Model. While I shall briefly describe some topics covered, 
a ``review of reviews'' is in no way a substitute for the original reviews
which also appear, of course, in this same volume. 

Amongst the more traditional topics I will cover are:  BFKL physics - 
higher-order corrections and jet cross-sections; unitarity and eikonal 
screening - mainly in deep-inelastic diffraction but also in soft diffraction; 
elastic scattering phenomenology - including real parts, the pomeron 
intercept and small-$t$ oscillations. I will also discuss the role 
of ``factorization'', i.e. both Regge pole factorization and perturbative 
QCD factorization theorems in the definition of a pomeron structure 
function and in the formulation of a ``parton model'' description of 
diffractive hard physics. I will focus on ``one gluon versus two gluons'' 
as illustrating the issues involved.

\section{QCD on the Lattice}

The ``coming of age'' of lattice gauge theory, as he called it, was reviewed
by G. Kilcup. 
This maturity is an outcome of many years of improving
techniques, algorithms 
etc. Particularly notable is the relatively recent improvement of 
lattice perturbation theory via tadpole summation. This process leads to the
scaling of the Wilson link variable by $U_{\mu} \longrightarrow U_{\mu}/ U_0 $
where $U_0$ is the mean-field value. The consequences have included
impressive results for the $\Upsilon$ spectrum and spin-splittings. A very
accurate value for $\alpha_s$ is obtained 
$$ 
\alpha_{\centerunder{---}{$\scriptstyle{MS}$ }}(M_Z) 
~=~ 0.1174 ~(\pm 0.0024) 
$$
Since this is more accurate than any other existing ``measurement'' it 
surely focusses attention on the assumptions made in refining lattice QCD 
to this level.  

Other new results include values for the strange quark mass.
Wilson and staggered fermions give different results, i.e. 
$$
m_s(\raisebox{2mm}{\centerunder{---}{$\scriptstyle{MS}$ }},2 GeV) 
= 100 \pm 21 MeV ~~ \hbox{and} ~~68 \pm 12 MeV
$$
respectively. Clearly there is a remaining 
need to improve lattice perturbation theory for fermions. Currently a number 
of groups are working on this problem and several ideas have been proposed,
although no method has yet been universally adopted. Kilcup
also presented new results for the glueball spectrum, including 
$$ 
M_{0^{++}} ~= 1630 \pm 60 \pm 80 MeV,~~~ M_{2^{++}}~= 2400 \pm 10 
\pm 120 MeV
$$

\section{Recent Advances in Perturbative QCD Calculations}

Perturbative QCD is now applied to a wide range of high-energy jet, photon,
heavy quark and weak vector boson production processes. Such processes also
inevitably provide the background in searches for new physics and so a detailed
understanding of the ``known physics'' involved is crucial. This 
is, perhaps, most clearly the case for higher-order contributions to jet
physics. As D. Kosower reviewed, the calculation of next-to-leading
order QCD contributions has become 
a forefront theory industry. At next-to-leading order and beyond,
sophisticated techniques are essential just to handle the enormous number of
diagrams, the large amount of vertex algebra in each diagram and the 
complexity of loop integrals with large powers of the loop momentum in the 
numerator. ``String-based rules'' for
organizing diagrams etc. have been particularly successful. The spinor
helicity method and supersymmetry decompositions have also been 
used very effectively.

Most recently, the string-based  rules have been superseded by
``unitarity-based'' rules. ``Cut-containing'' parts of amplitudes are
determined by sewing together on-shell tree amplitudes via unitarity
integrals. An explicit dispersion relation is not used, instead the sewing 
procedure is used only to determine the integral functions that can appear 
in a given amplitude, along with their coefficients. The restricted number 
of functions that can appear actually implies that in some cases cut-free
pieces are also determined by the sewing process. In supersymmetric 
theories, for example, entire amplitudes are determined. In general, 
the remaining (real part) ambiguity of possible ``rational
pieces'' is determined by collinear limits and/or dimensional
regularization. Kosower summarized, by noting that a variety of
hard calculations have been completed using these techniques and ``two loops
is the next frontier''. 

\section{Effective Field Theories in Strong Interactions}

Effective field theories are constructed by combining symmetry 
considerations with a power counting algorithm relevant to the kinematic regime 
studied. The subject was reviewed by M.~ Wise.

The first example considered 
was chiral perturbation theory for nucleon-nucleon interactions. Pion 
exchange is the lowest-order interaction, but a problem arises. Iteration of 
this interaction leads to ultra-violet divergences which generate a 
leading-order counter term which should be suppressed according
to the power counting algorithm used. 
As a result the consistency of the formalism is in doubt.

The next effective field theory reviewed was NRQCD. Here an expansion in
powers of ``$v/c$'' gives surprising predictions for quarkonium
production/decay because suppressed powers of $v/c$ can be enhanced by
factors of $1/\alpha_s$. The separation of color singlet and octet
contributions via an operator product expansion has to be treated with
particular care when ``end-point'' higher-order contributions are involved.
In general a careful study of end-point phenomena provides an understanding
of the failures of ``octet production''.  Successes of the ``octet
mechanism'' were described in separate talks by G. Zhao and J.
Lee. 

The third example discussed by Wise was heavy quark effective theory. In this 
case the limit $m_Q \to \infty$, with $v$ fixed, breaks both spin and flavor 
symmetries. The remaining light quark spin symmetry can be used to obtain 
mass differences, sum rules, etc. for heavy quark hadrons. Such sum rules
were discussed in detail by M.~Bander. In particular Bander noted
that a sum rule for heavy meson decay widths works well for charmed quark
mesons and even kaons can be successfully treated as heavy quark hadrons! 

\section{ Other QCD Topics}

Other ``non-diffractive'' QCD talks included the following. 

\begin{itemize}

\item {A very interesting 
all-orders discussion of ultra-violet renormalons in QED (with some 
discussion of QCD) 
was presented by T.~Lee. He considered the effective charge and
showed that, at all orders, the strength of the singularity is independent
of the number of exchanges. } 

\item{ C.~Ji suggested that a factorization prescription 
provided by light-cone quantization might allow perturbative QCD to be used
to study exclusive processes at intermediate energies. }

\item{ Techniques developed for studying the nucleon-nucleon force were 
reviewed by R.~Vinh Mau.}

\item{ Features of multiplicity distributions, including 
the kinds of pattern recognition that can be used, were reviewed by 
I.~Dremin.}

\item{ T.~Muta
described the four loop calculation of the QCD $\beta$-function 
}

\item{ Polarized deep-inelastic scattering was discussed by both 
Muta and S. Troshin. }

\item{ Low-energy form factors were discussed by W.~Buck. }

\item{ Inclusive B meson decays 
were considered by Y.~Keum while semileptonic B decays 
were discussed by D.~Hwang.}

\end{itemize}

\section{The Current Status of Quantum Electrodynamics}

T. Kinoshita described how low-energy, high precision, experiments now 
provide very accurate measurements of the anomalous magnetic moments of the 
electron and the muon, as well as the hyperfine structure of muonium. 
He pointed out that the ``best measurement'' of $\alpha_{em}$ is 
obtained by demanding the theoretical consistency  of measurements of
the anomalous magnetic moment of the electron. The result is 
$$
\alpha_{em}(a_e) = 137.03599993(52)
$$
which is an accuracy of $3.8 \times 10^{-9}$. Soon $\alpha_{em}$ will be 
directly measured to an accuracy better than $10^{-9}$ and the question of 
whether this fundamental constant is truly universal will be become a 
significant issue. 

\section{Beyond the Standard Model}

Possible deviations from the Standard Model were the focus of several talks.

\subsection{SUSY at the Tevatron}

 C. S. Kim considered whether the excess jet cross-section 
seen~\cite{CDF} by CDF at large $E_T$ can be explained by virtual SUSY effects.
After including one-loop corrections in the running of $\alpha_s$ and
the appropriate modification of parton distributions, Kim concludes that the 
the CDF effect is too large to be a SUSY effect.

\subsection{Large $x$ and $Q^2$ at HERA}

The large $x$ and $Q^2$ events~\cite{dav}
seen at HERA were discussed by P. Ko.
He argued that if R-parity violation is allowed  within the minimal
supersymmetric standard model, these events can be interpreted as 
$s$-channel ``stop'' production. 

\subsection{Model Construction}

General strategies for extending the Standard Model were reviewed by P. 
Frampton. He emphasized the need for motivation and testability. As 
illustrative examples he considered the Left-Right model as 
explaining the chirality of quarks and leptons, the ``331'' model as 
providing an understanding of the existence of three generations, and the
SU(15) model as an example which accomodates light leptoquarks without
producing proton decay.

\section{BFKL Physics}

\subsection{NLO and the Effective Action}

The completion of the calculation of next-to-leading-order 
corrections to the BFKL kernel was described by L. Lipatov. 
There are three components of the kernel -

[1] the Regge trajectory of the gluon, [2] the reggeon-reggeon-particle 
vertex and
\newline \indent [3] the four-reggeon interaction.

\noindent NLO corrections to [1] and [2] have been known for several years. The 
corrections to [3] arise from the production of pairs of gluons (and quarks)
separated by a finite rapidity gap. They are very complicated and have taken
a long time to calculate. 

Lipatov further described how a gauge-invariant effective lagrangian can be 
constructed for gluon-reggeon interactions. The production of finite 
rapidity multi-gluon states can be included in the effective 
lagrangian, although the NLO results have not yet been derived this way. 
Additional properties of multi-reggeon states, including the odderon, were 
also discussed. 

There is no doubt that the NLO calculation is a historic contribution to 
understanding small-$x$ and Regge limit physics in QCD. Not surprisingly,
many important questions remain, including the following, some of which will
surely be answered soon. 

\noindent 1) What is the new BFKL pomeron intercept, i.e. the new 
leading eigenvalue?
\newline 2) How does the scale that appears in $\alpha_s$
contribute to small-$x$ evolution? 
\newline 3) What do the NLO corrections tell us
about the kinematic range of validity of 
\newline \indent BFKL physics. 
\newline 4) Is there any role for conformal symmetry at NLO? In particular, 
what is the relationship 
\newline \indent the NLO kernel obtained~\cite{cww} via $t$-channel 
unitarity which, in impact parameter space, is 
\newline \indent the fourth power of a 
logarithm of a simple harmonic ratio?

\subsection{Jet Physics}

BFKL jet production was discussed by V. Kim. When considering dijet 
production at large rapidity separation, there is an issue as to whether 
BFKL effects should be included in the parton distributions. Is this double 
counting? Kim argued that including BFKL in the parton distributions used
could explain the small $x_T$ anomaly found by CDF (when data at
$\sqrt{s}= 630 $ GeV and $\sqrt{s}= 1800 $ GeV are compared).

\section{Eikonal Screening and Unitarity Corrections}

The eikonal model is commonly used as an approximation to $s$-channel 
unitarity. It was used by several speakers to discuss screening in DIS
diffraction and vector meson production. 

\subsection{Vector Meson Production}

U. Maor used a two radii model of the proton to discuss screening 
corrections to diffractive vector meson production. The values of the radii 
were chosen to fit the data for $\psi$ diffractive production. Amongst the 
conclusions are 

\begin{itemize}

\item[{1)}] $\alpha_{eff}'$ decreases with $Q^2$

\item[{2)}] a dip in ${d \sigma \over d t }$ is predicted

\item[{3)}] screening corrections decrease the increase with $Q^2$ of the 
pomeron intercept.

\end{itemize}

Eikonal screening and unitarity, in the context of the general issue of
separating hard and soft diffraction, was discussed by V.~Petrov. He argued 
that the HERA data on vector meson production can be fit without the 
addition of any ``hard pomeron'' trajectory. The approach to asymptopia is 
simply delayed at larger $Q^2$.

\subsection{Unitarity Limits In Impact Parameter Space}

It was argued by E. Predazzi that diffractive diisociation of a highly 
virtual photon is predominantly a soft process and that unitarity effects 
should be as important as in hadronic reactions.  Predazzi proposed an
impact parameter analysis to detect the onset of ``unitarity effects'' 
in the data. The expected effects are most pronounced when alaysed this way.
In particular, $\Delta_{eff}(b)$ is predicted to be significantly reduced at
small $b$. 

When the energy and impact parameter dependence of the 
eikonal are factorized, the increase of the cross-section with energy is 
directly felt at all impact parameters. As a result the unitarity limit is 
rapidly approached. As was discussed by M.~Gay Ducatti, vector meson
production at HERA is actually well-fit by such a model. In this case, the
unitarity bound is indeed being rapidly approached. 

\subsection{Diffractive Phenomenology}

Unitarity corrections to the large mass diffractive cross-section
(i.e. the ``triple-pomeron'' cross-section) were discussed by C.~Tan. 
If the whole cross-section is fit to the triple pomeron formula with
$\alpha_{\spom} \hspace{0.5mm}(0) > 1$ then unitarity problems with energy 
dependence quickly appear. Eikonal 
screening can be employed to avoid this and can, perhaps, be used as a
basis for ``pomeron flux renormalization'' phenomenology. However, since 
very strong screening corrections are invoked, the Regge pole factorization
property is lost and, as we discuss further below, the basis for a parton
model analysis of the pomeron is also lost. 

The eikonal model does not include ``unitarity corrections'' which 
occur over partial (finite) rapidity ranges away from the ends of the rapidity 
interval. A simple example of such processes is multiple diffraction. 
Another example is the appearance of heavy quark states. Tan refers 
generically to all such processes as ``flavoring''. Flavoring can be 
incorporated phenomenologically by allowing the pomeron intercept to be 
rapidity-interval dependent. A successful phenomenology can be developed 
with the Regge pole factorization property of the pomeron retained. 

A full phenomenological analysis of diffraction employing secondary Regge
trajectories has not yet been carried out (although H1 do subtract the
contribution of secondary trajectories when extracting the diffractive 
cross-section~\cite{h1}). In fact 
one would expect comparable phenomenological success to that obtained via 
flavoring, since including secondary Regge trajectories will produce 
an effective rapidity-dependent intercept. 

\section{Elastic Scattering Phenomenology}

\subsection{The Pomeron Intercept}

The soft pomeron intercept $\alpha_{\spom}\hspace{0.5mm}(0)$ plays an important role in 
HERA diffractive phenomenology. Is the well-known Donnachie-Landshoff result
the best fit to all high-energy data? This question was discussed by K.~Kang

Inconsistencies amongst data sets suggest that a ``filtering'' process 
should be applied. In this process the complete set of all data points is 
fit first and those data points which are more than $2 \sigma$ away from the 
fit determined. These data points are removed and the fit is again 
performed. This time the data points that are more than  $1 \sigma$ away 
from the fit are removed. Finally the remaining data points are refit. The 
vital point is to determine whether the parameters of the fit remain stable 
during the filtering process.

Kang argued that the Donnachie-Landshoff fit, with exchange-degenerate 
non-leading trajectories, is unstable under the filtering process. If 
exchange degeneracy is not imposed, a fit is obtained that is stable under 
filtering. The resulting pomeron intercept is
$$
\alpha_{\spom}\hspace{0.5mm}(0)~=~1.0964 ~~~+0.0115 /-0.0091
$$
which is somewhat higher than the $1.08$ of the Donnachie-Landshoff fit.

A phenomenology based on the use of two pomeron poles, as an approximation to 
the BFKL pomeron, was described by L.~Dakhno. This leads to 
$$
\alpha^{(1)}_{\spom}~=~1.29~, ~~~~~\alpha^{(2)}_{\spom}~=~1.00
$$
giving a better fit to the BFKL ``hard pomeron'' intercept. 

\subsection{The Real Part} 

The UA4/2 results for small $t$ scattering, and the related result
for the real part of the hadronic amplitude, continue to provoke discussion.
To extract the real part it is essential that the hadronic amplitude can be
parametrized by a simple exponential. 

Both P.~Gauron and O.~ Selyugin argued that the UA4/2 results suggest the 
differential cross-section actually has very small scale oscillations on top
of the exponential behavior. This would, of course, invalidate the
conventional extraction of the real part. Gauron suggested such oscillations
could be a manifestation of the Auberson-Kinoshita-Martin zeroes that are
required by general asymptotic theorems. In conventional models of the
elastic scattering diffraction peak the AKM zeroes are simply manifest in
the familiar diffraction pattern of dips and maxima which are established
experimentally. These represent ``oscillations'' which are present with a
period set by normal hadronic scales. If the oscillations suggested by
Gauron and Selyugin are indeed present they indicate the exsatence of a
completely new hadronic scale. The period in $\sqrt{|t|}~$ is $\approx ~20$
MeV. 

A related subject was discussed by M.~Block. He considered the real part 
zero that A.~Martin has recently argued should be present in models of the 
kind conventionally used to fit current data. In the ``QCD inspired'' eikonal 
model discussed by Block this zero actually appears at much larger $|t|$ and
is, as would surely be generally expected, simply related to the normal
diffraction pattern. 

Block also noted that the anomalous magnetic moment of the nucleon is always 
ignored when the real part is extracted via Coulomb interference with the 
electromagnetic interaction. He estimated that when this effect is taken 
into account the $\bar{p}p$ real part is lowered by $\approx ~0.0075$ 
while the $pp$ real part should be raised by the same amount. This shift is 
significant since it is comparable to the error quoted by UA4/2 on their 
result. Regarded as a theoretical uncertainty, the effect can be minimized 
by splitting the interference region into two parts and extracting 
two experimentally independent results.

\section{Factorization and the Pomeron Structure Function}

Finally, I would like to discuss a central theoretical issue which is
now being confronted experimentally.  Can the QCD parton model be extended
to the pomeron? 
The simplest possibility, which comparison of HERA and Tevatron 
data already seems to rule out~\cite{ma}, would be that the pomeron is like a
hadron and has a universal structure function that can be measured in
DIS and transported to hadron hard diffraction as illustrated in Fig.~1.
In fact, since the pomeron is a virtual exchange, we expect it to be
different to a normal hadron. To discuss this we first distinguish two
basic ``factorization'' concepts that are involved. 

\noindent \parbox{3.9in}{
\begin{center}
\leavevmode
\epsfxsize=3.4in
\epsffile{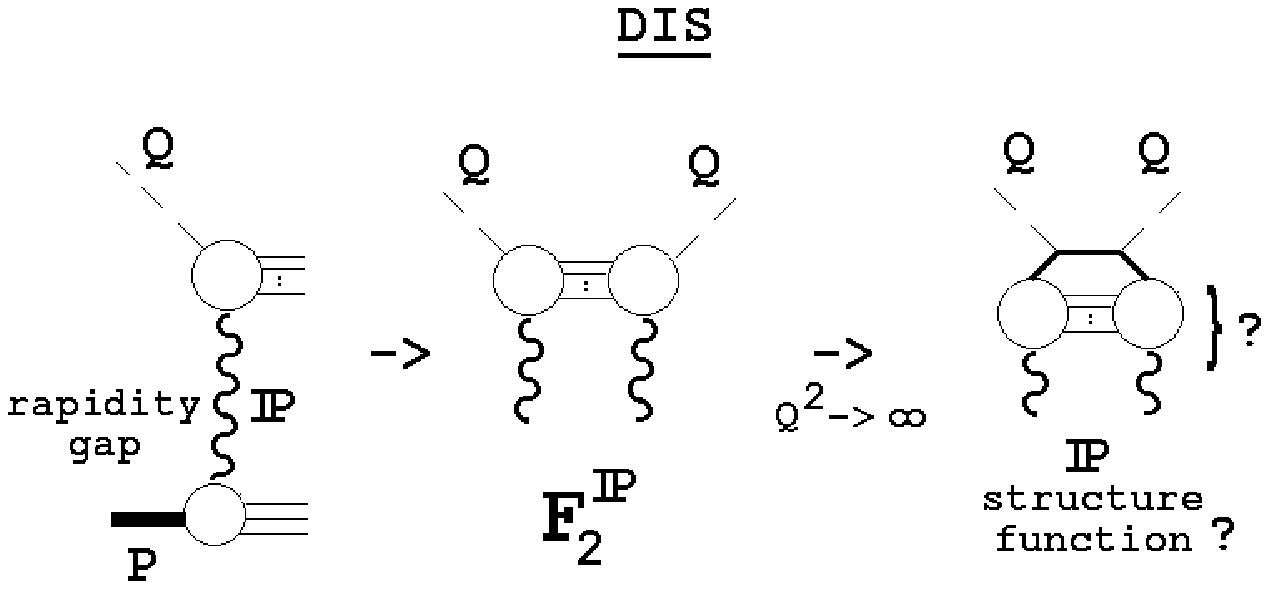}
\end{center}
}
\parbox{2.1in}{
\begin{center}
\leavevmode
\epsfxsize=1.5in
\epsffile{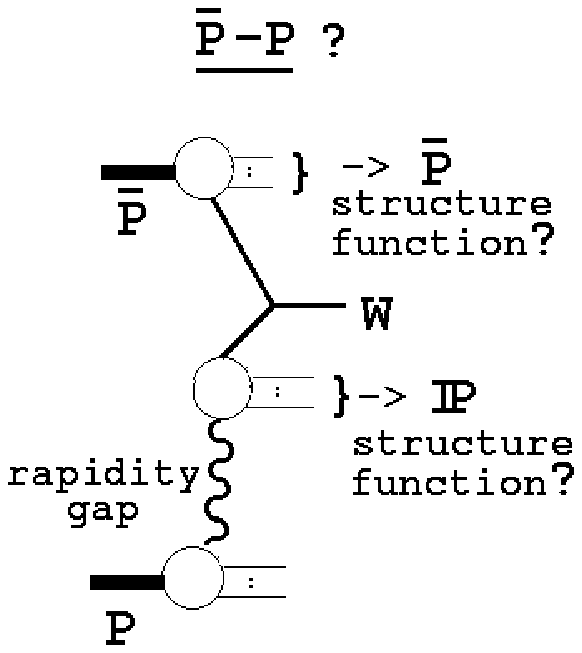}
\end{center}
}
\newline \centerline{Fig.~1}

\subsection{Regge Pole Factorization}

This follows theoretically from the $J$-plane continuation of $t$-channel 
unitarity discussed in my talk. If the pomeron is (approximately) a Regge
pole, the factorization of residues ensures 
that the first stage of Fig.~1 giving $F^{\spom}_2$ is well-defined.
(Experimentally it is straightforward to subtract the contribution of 
secondary reggeons.)

It is not generally appreciated, but that there are two fundamental theoretical
reasons why the pomeron should be a Regge pole. Firstly, multi-pomeron
$t$-channel states are well-defined and multiparticle $t$-channel unitarity 
(i.e. reggeon unitarity) is satisfied only if the pomeron is a Regge pole.
More physically, perhaps, if the parton model has it's origin in infinite
momentum quantization (as we discuss below) the pomeron will be a direct
manifestation of the ``universal wee parton distribution'' in a hadron. 
Wee parton universality requires Regge pole factorization for the 
pomeron~\cite{arw}. 

\subsection{Factorization Theorems and the ``QCD Parton Model''}

Factorization theorems provide a basis for the application of
perturbative QCD. If $Q^2$, say, is a large scale such that $\alpha_s(Q^2)$ is
small, perturbation theory can be consistently used 
provided any infra-red divergences that appear can be factorized into
parton distributions. This leads to the familiar ``QCD parton model'' for
inclusive production, i.e. (symbolically) 
$\sigma ~ \sim ~ \int f_a f_b ~\sigma_{ab}$
where $\sigma_{ab}$ is a parton cross-section and $f_a$ and $f_b$ are
parton distributions. Applying the renormalization 
group to the factorization process leads to an evolution equation for the 
parton distributions. Note that factorization theorems, and therefore the
QCD parton model, apply only to the leading power in $Q^2$ (leading-twist). 

For Fig.~1 to hold in it's entirety, a QCD factorization theorem must be 
valid at large $Q^2$ when $F^{\spom}_2$ is defined by Regge factorization.
At present there is no indication that both factorization properties can 
be simultaneously satisfied within QCD. Perturbatively, the pomeron 
appears as two (reggeized) gluon exchange and is not a Regge pole.
Alternatively, one would not expect that a conventional factorization proof
applies for the full ``off-shell'' non-perturbative pomeron.

Regge pole factorization can, perhaps, be by-passed by defining~\cite{vt}
an inclusive 
DIS cross-section in which the initial proton is required to go forward 
(if there are no massless particles a rapidity gap must appear at
high-energy). At large $Q^2$ this cross-section may satisfy a factorization 
theorem and provide a perturbative basis for describing some
properties of DIS diffraction. However, since the lowest-order perturbative
contribution to the pomeron (i.e. the rapidity gap cross-section) would be
two gluon exchange, it is 
apparent phenomenologically (from the scaling violations in 
particular) that non-leading twist terms must be equally important
in the current experimental cross-section~\cite{bw}. As a result, it seems
that this formulation of a leading-twist ``QCD parton model'' for
diffraction can at best be 
relevant experimentally at much higher $Q^2$. (Note that it is also unlikely
that the formalism can be extended to hadron scattering.) 

\subsection{The Infinite Momentum Frame Parton Model}

There are strong indications that in processes involving final state 
hadrons, where a-priori a QCD factorization theorem is much more difficult to 
prove, some form of the parton model is valid significantly beyond the 
leading-twist approximation. Most notable is the
success of  dimensional counting rules in elastic scattering~\cite{az}
and even, perhaps, the constituent quark model itself~\cite{kw}. If a
broader form of the parton model is valid in QCD we may expect, as
we now discuss, that it will be directly visible in the parton properties of
the pomeron. 

The original parton model formulated by Feynman, Gribov and others, 
was indeed a broader concept, in many respects, than the ``QCD parton model''
we have described above. It was based on the (superficial) simplicities of 
infinite momentum frame quantization. At infinite momentum the constituents
of a theory are apparently exposed directly without the complications of
the vacuum being present. If the vacuum is non-trivial, as it is of course
in QCD, the ``wee partons'' (with vanishingly small momentum fraction) must 
somehow carry the vacuum properties of the theory, including confinement 
and chiral symmetry breaking. This is a very strong
requirement which may well not be satisfied in QCD. To carry vacuum
properties, the wee partons must certainly be the same in all hadrons
(Feynman proposed that the wee distribution should be that of a critical
phenomenon). If the pomeron is produced by the interactions of universal wee
partons it will also be universal and so will have the regge pole
factorization property~\cite{arw}. 

In my talk I discussed breaking the SU(3) gauge symmetry of QCD to SU(2).
In this case, the Regge pole nature of the pomeron is directly related to the
presence of a universal infra-red divergent component of both 
hadrons and the pomeron.  A ``reggeon condensate'' is produced which is
responsible for hadron vacuum properties and, in a sense, is the simplest
possible universal wee-parton distribution. If the full gauge symmetry is 
restored the wee parton distribution (when it is independent of the 
$k_{\perp}$ cut-off) is that of the Critical Pomeron.
I described how the special properties of the wee parton
component lead to the single gluon dominance of the DIS pomeron structure
function seen in the H1 analysis~\cite{h1}. 
\newline \parbox{3.8in}{\openup\jot Single gluon dominance 
will also occur in hadronic hard diffraction. However, the 
wee parton component of the pomeron plays an even more crucial role. 
As illustrated in Fig.~2, 
this component must couple to the hadron state whose hard 
constituent produces the $W$. Thus a new hadron structure function
is involved, distinct from that appearing in normal parton 
model processes.
In this sense, the $\bar{p}p$ part
of Fig.~1 does not 
hold. However, a more subtle parton model may nevertheless be at work.}
\parbox{0.2in}{$~$}
\parbox{2.1in}{
\begin{center}
\leavevmode
\epsfxsize=1.7in
\epsffile{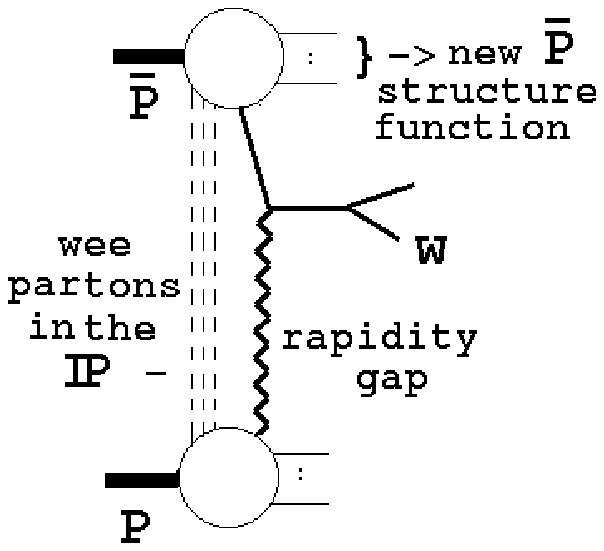}

Fig.~2
\end{center} 
}

It is my belief 
that the picture of the pomeron I presented will eventually be understood as
signifying the validity of a deeper form of the parton 
model in QCD, beyond that of leading-twist factorization theorems. The 
immediate issue is whether a two-gluon or single gluon picture of the hard 
component of the pomeron is most successful experimentally.

\section*{References}

Explicit references are given only to topics not directly covered in a 
workshop talk.


\begin{thebibliography}{99}

\bibitem{CDF} CDF Collaboration, {\it Phys. Rev. Lett.} {\bf 77}, 438 
(1996).

\bibitem{dav} M.~David - these proceedings.

\bibitem{cww} C.~Corian\`{o}, A.~R.~White and M.~W\"usthoff,
{\it Nucl. Phys.} {\bf B493}, 397 (1997);
C.~Corian\`{o} and A.~R.~White , {\it Nucl. Phys.} {\bf B468}, 
175 (1996), {\it Nucl. Phys.} {\bf B451}, 231 (1995).

\bibitem{h1} H1 Collaboration, pa02-61 ICHEP'96 (1996). For a final version
of this analysis see hep-ex/9708016 (1997). 

\bibitem{arw} A.~R.~White, {\it Phys. Rev.} {\bf D 29}, 1435 (1984).

\bibitem{ma} M.~Albrow - these proceedings.

\bibitem{vt} G.~Veneziano and L.~Trentadue, {\it Phys. Lett.} {\bf B323}, 
201 (1994); A.~Berera and D.~Soper, {\it Phys. Rev.} {\bf D 50}, 4328 (1994).
  
\bibitem{bw} J.~Bartels and M.~W\"usthoff, Contribution to DIS97 - 
ANL-HEP-CP-97-51 (1997)

\bibitem{az} A.~R.~Zhitnitsky, hep-ph/9605226 (1996).

\bibitem{kw} K.~G.~Wilson, T.~S.~Walhout, A.~Harindranath, Wei-Min Zhang, 
S.~D.~Glazek and R.~J.~Perry, {\it Phys. Rev.} {\bf D 49}, 6720 (1994).

\end{thebibliography}
\end{document}